\documentclass[conference]{IEEEtran}

\usepackage{blindtext}
\usepackage{eso-pic}
\IEEEoverridecommandlockouts
\usepackage{cite}
\usepackage{amsmath,amssymb,amsfonts}
\usepackage{algorithm}
\usepackage{algpseudocode}
\usepackage{amsthm}
\usepackage{multicol}
\usepackage{pslatex}
\usepackage{graphicx}
\usepackage{textcomp}
\usepackage{xcolor}

\usepackage{eso-pic}
\newcommand\AtPageUpperMyright[1]{\AtPageUpperLeft{%
 \put(\LenToUnit{0.17\paperwidth},\LenToUnit{-2cm}){%
     \parbox{0.9\textwidth}{\raggedleft\fontsize{8}{11}\selectfont #1}}%
 }}%
\newcommand{\conf}[1]{%
\AddToShipoutPictureBG*{%
\AtPageUpperMyright{#1}
}
}

\begin{document}
\title{\vspace*{1cm} A Hybrid Approach for Smart Alert Generation
}

\author{\IEEEauthorblockN{Yao Zhao}
\IEEEauthorblockA{\textit{Cisco Meraki} \\
San Francisco, USA \\
yaozhao2@cisco.com}
\and
\IEEEauthorblockN{Sophine Zhang}
\IEEEauthorblockA{\textit{Cisco Meraki} \\
San Francisco, USA \\
sophizha@cisco.com}
\and
\IEEEauthorblockN{Zhiyuan Yao}
\IEEEauthorblockA{\textit{Cisco Meraki} \\
Paris, France \\
yzhiyuan@cisco.com}
}

\maketitle
\conf{\textit{  Proc. of the International Conference on Electrical, Computer, Communications and Mechatronics Engineering (ICECCME 2023) \\ 
19-20 July 2023, Tenerife, Canary Islands, Spain}}
\begin{abstract}
Anomaly detection is an important task in network management.
However, deploying intelligent alert systems in real-world large-scale networking systems is challenging when we take into account (i) scalability, (ii) data heterogeneity, and (iii) generalizability and maintainability.
In this paper, we propose a hybrid model for an alert system that combines statistical models with a whitelist mechanism to tackle these challenges and reduce false positive alerts.
The statistical models take advantage of a large database to detect anomalies in time-series data, while the whitelist filters out persistently alerted nodes to further reduce false positives. Our model is validated using qualitative data from customer support cases. Future work includes more feature engineering and input data, as well as including human feedback in the model development process.
\end{abstract}

\begin{IEEEkeywords}
Anomaly Detection, Statistical Model, Alert System
\end{IEEEkeywords}

\section{\uppercase{Introduction}}
\label{sec:introduction}

Network alert and anomaly detection systems are essential for predicting and preventing potential issues in networking systems~\cite{lad2006phas}.
To ensure seamless operations, large organizations have developed their own anomaly detection services to monitor their products and services, which aim to detect anomalies and raise alerts for timely decision-making related to incidents~\cite{laptev2015generic,vallis2014novel}. For example, Yahoo has developed EGADS~\cite{laptev2015generic}, which automatically monitors and generates alerts for millions of time-series data related to different Yahoo properties. Microsoft also utilizes an anomaly detection service to monitor millions of metrics from Bing, Office, and Azure, which has enabled engineers to quickly address live site issues~\cite{ren2019time}.

Providing too many alerts can be overwhelming for customers and negatively impact the quality of service, while missing critical events can result in delayed reaction to incidents~\cite{hussain2019artificial}.
Yet, it is challenging to develop an intelligent and effective alert system, that can accurately distinguish exceptional events that have the potential to lead to networking issues, especially in large-scale systems with a high volume of events. 

\textbf{Challenge 1: Scalability.} With the huge amount of data generated by networking systems, it is critical to use efficient algorithms to process this data.
In addition, the system must be able to handle the large number of devices and customers that it serves.
Deep-learning based approaches demonstrate promising results, yet they incur significant overhead when deployed for large scale systems~\cite{hussain2019mobile, yin2022tsmc}.
Our proposed approach addresses this challenge by using a hybrid model that combines both a statistical model and a rule-based whitelist mechanism.

\textbf{Challenge 2: Data Heterogeneity.}
While efficient mechanisms have been proposed to extract observations from the data plane and help analyze system states~\cite{yao2022efficient}, it is intrinsically hard to determine if a networking issue has actually occurred, as there may be multiple factors that contribute to an event~\cite{ren2019time}.
To address this challenge, we incorporate qualitative data, \textit{i.e.} customer support cases, to provide a more comprehensive understanding of the data.
This allows us to better identify exceptional events and improve the accuracy of the system.

\textbf{Challenge 3: Generalizability and Maintainability.}
As the system is deployed to millions of networking devices and millions of customers for a variety of alerts, it is essential to take into account multiple objectives, such as alert accuracy, reliability, and most importantly, generalizability and maintainability.
Both the statistical model and the additional whitelist mechanism in our proposed approach can be generalized and applied for a wide range of alert generations.
Our approach also allows incorporating human input to improve the reliability of the data-driven system and to enable iterative model development.

\subsection{Related Works}

Various anomaly detection algorithms have been proposed in the literature, including supervised learning, unsupervised learning, and statistical approaches.

To improve the accuracy of anomaly detection, supervised models have been investigated. For instance, EGADS~\cite{laptev2015generic} used a collection of anomaly detection and forecasting models along with an anomaly filtering layer to enable scalable anomaly detection on time-series data. Opprentice~\cite{opprentice} achieved superior performance compared to traditional detectors by utilizing statistical detectors as feature extractors and detecting outliers with a Random Forest classifier.
However, supervised approaches are insufficient in online applications since continuous labels cannot be obtained in industrial environments. However, the networking domain poses unique challenges due to the absence of ground truth label data and the variability of data. Supervised learning methods require labeled data, which is difficult to obtain proactively until customers report connectivity issues.

To tackle these problems in industrial applications, unsupervised approaches have been studied. DONUT~\cite{donut} is an unsupervised anomaly detection method based on Variational Auto-Encoder (VAE), which models the reconstruction probabilities of normal time-series. Abnormal data points are reported if the reconstruction error was larger than a threshold.
Luminol~\cite{luminol} computes anomaly scores by segmenting time-series into chunks and evaluating the frequency of similar chunks.
However, unsupervised learning algorithms may lack interpretability, making it difficult for engineers to understand the results and take appropriate actions.

As a result, our model is based on statistical algorithms that are simple and efficient, especially for large-scale network databases. 
A variety of statistical models have been investigated, such as hypothesis testing~\cite{hypothesis_testing}, wavelet~\cite{wavelet}, singular value decomposition (SVD)~\cite{svd}, auto-regressive integrated moving average (ARIMA)~\cite{arima}, and Fast Fourier Transform (FFT)~\cite{fft}. For instance, FFT helps identify time-series segments with high frequency changes, which can be verified with the Z-value test.
These algorithms provide interpretable results and insights into the system characteristics, which is crucial for early-stage development when engineers need to understand the context.

\begin{figure}[!t]
  \centering
    \includegraphics[width=\columnwidth]{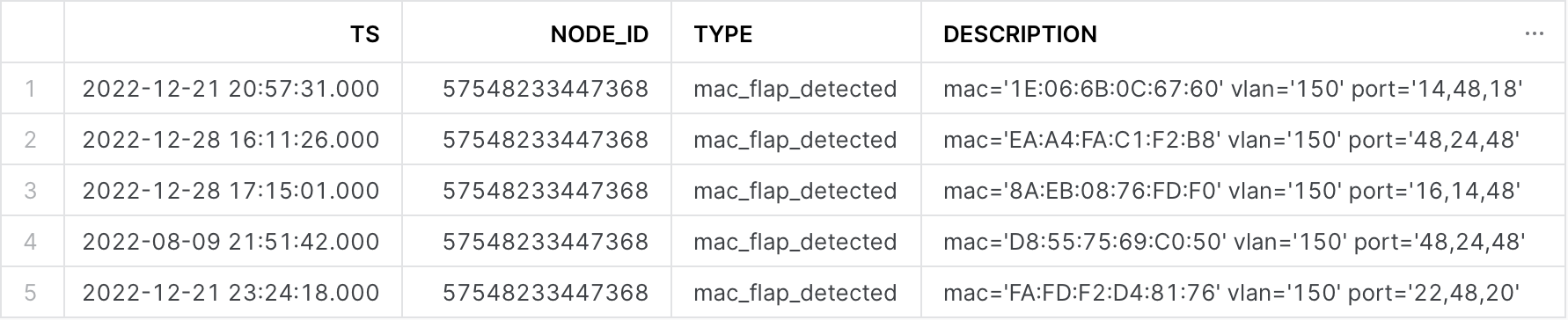}
    \vskip -.1in
    \caption{MAC-flap event log.}
  \label{fig:bg_mac_flap}
  \vskip -.1in
 \end{figure}

\begin{figure}[!t]
  \centering
    \includegraphics[width=\columnwidth]{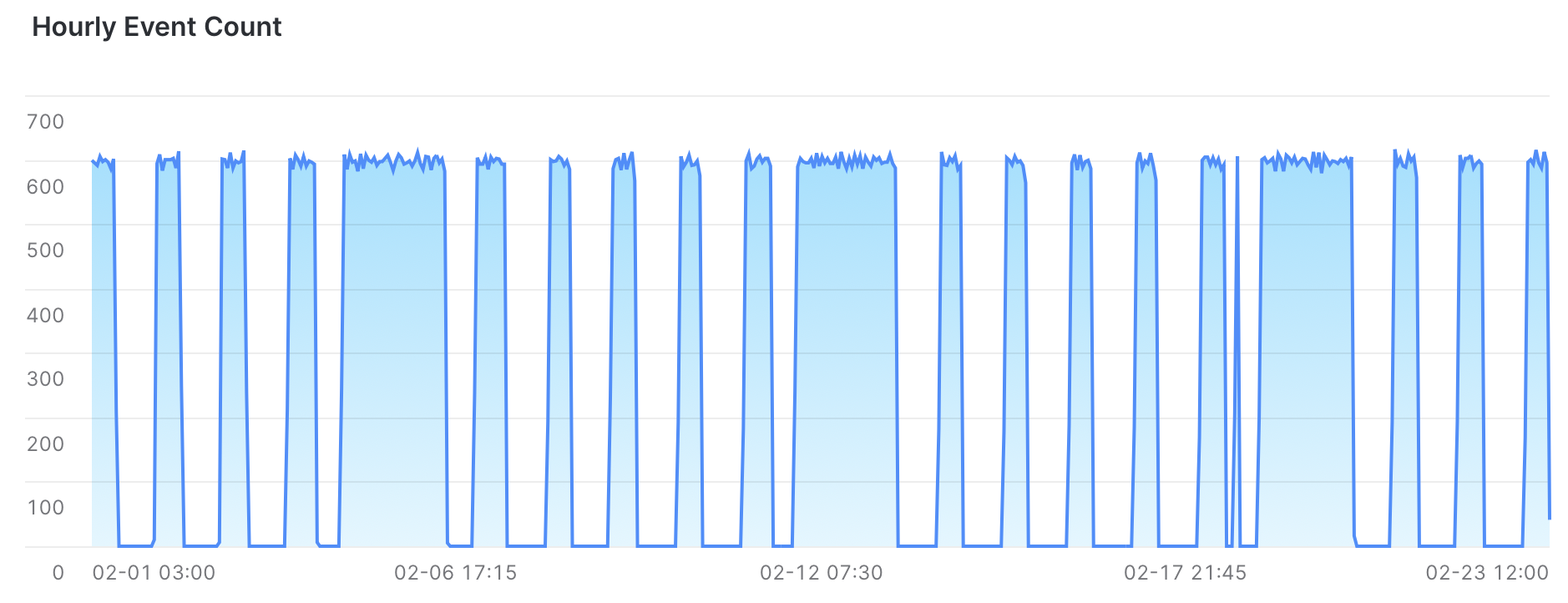}
    \vskip -.1in
    \caption{1 hour aggregated MAC-flap event count.}
  \label{fig:1h_aggr}
  \vskip -.1in
\end{figure}

\subsection{Statement of Purpose}

In this paper, we propose a hybrid approach for building a smart alert system that balances multiple objectives, including alert accuracy, alert ratio, alert reliability, and model maintainability. Our approach takes multiple telemetries into account and incorporates both quantitative and qualitative data feedback from humans. We present a novel statistical model that integrates multiple data sources to generate alerts for exceptional events. We also introduce a framework for developing and deploying such systems that takes into account trade-offs among multiple objectives. The proposed framework enables us to operate the model at scale and upgrade it with more user feedback.

The main contribution of this paper is the proposed hybrid approach that leverages both quantitative and qualitative data for building a smart alert system that accurately identifies exceptional events in networking systems. Our approach not only addresses the challenge of building an accurate alert system, but also provides a framework for balancing multiple objectives when developing and deploying such systems. 
We believe that our proposed approach can have significant implications for the networking industry, and we present our experimental results to demonstrate the effectiveness of our approach.

\section{\uppercase{Overview}}
 
Network anomaly detection is critical for maintaining the health and stability of networking systems. One common use case in the networking domain is MAC-flap detection, which occurs when a Media Access Control (MAC) address is learned on multiple ports within a short period of time. MAC-flapping events can be persistent and hard to define the baseline pattern of on/off events. The Meraki switch has enabled MAC-flap detection as a default feature to monitor the MAC forwarding table and report flapping events to the dashboard, as depicted in Figure~\ref{fig:bg_mac_flap}.

\begin{figure}[!t]
  \centering
    \includegraphics[width=\columnwidth]{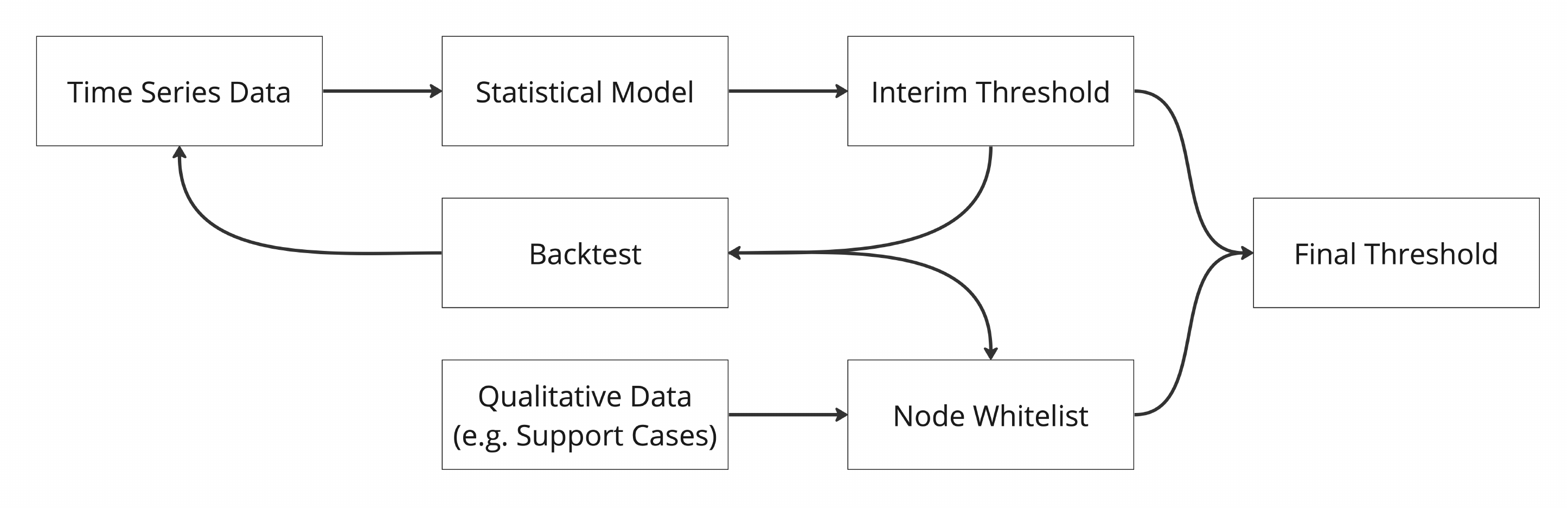}
    \vskip -.1in
    \caption{Threshold generation workflow overview.}
  \label{fig:overview}
  \vskip -.1in
 \end{figure}

Our model takes data sources generated on an hourly basis (as depicted in Figure~\ref{fig:1h_aggr}) and looks back at 5 weeks' worth of data to compute the statistical baseline for each node. Based on a threshold, the model checks if new data points are outliers. The overall process of generating the threshold is depicted in Figure~\ref{fig:overview}. An interim threshold is generated from the hourly aggregated time series data by the statistical model (see Section~\ref{sec:method_model}). Since Mac-flap events are sometimes normal, we also incorporate a whitelist created from back-tested $k$ weeks' alerts. This whitelist includes nodes that have received excessive alerts over multiple weeks, but no issues (\textit{i.e.} support cases) were reported by customers. By incorporating both statistical thresholds and a whitelist, our model improves the accuracy of Mac-flap detection and reduces false positives.

\section{\uppercase{Methodology}}
\label{sec:method}

Our proposed model for anomaly detection consists of two components - a statistical model and a whitelist mechanism.
The statistical model is responsible for identifying anomalies in the time-series data, while the whitelist acts as an additional filter to eliminate persistently alerted nodes that are not true anomalies.
By incorporating this additional mechanism, we aim to improve the accuracy and control of the alert system, particularly in enterprise-level products where false positive alerts can cause significant disruptions to the user experience.
In this paper, we provide a detailed description of our model and demonstrate its effectiveness in real-world applications.

\subsection{Statistical Model}
\label{sec:method_model}

Statistical methods rely on the assumption that anomalies are rare events that can be detected by deviations from the normal statistical distribution.
We chose statistical profiling for our use case because it is a simple and effective method of detecting anomalies in network organizations with millions of devices. 

Statistical profiling uses statistical analysis to create a profile of normal behavior.
In our use case, for each network, we calculated an upper boundary for MAC-flap event frequency using statistical analysis of past data.
We collected raw data on hourly MAC-flap event counts and calculated the week-over-week percent change on the hourly count.
The whole process is shown in algorithm~\ref{alg:statistical}.

\begin{algorithm}[t]
  \small
  \caption{MAC-Flap Smart Alert}\label{alg:statistical}
  \begin{algorithmic}
  \Require $g(d, h)$ \Comment{Get the time given the days to end date $d$ and hour $h$}
  \Require $f(t)$ \Comment{Get the hourly event numbers}
  \Ensure $f(t) \geq 0$
  \State $X \gets \left[\right]$ \Comment{Week-over-week percent change buffer}
  \State $Y \gets \left[\right]$ \Comment{Result}
  \State $N \gets 27$ \Comment{Time window size in day}
  \While{$N > 0$}
      \State $H \gets 24$ \Comment{Number of hours per day}
      \While {$H > 0$}
          \State $t \gets g(N, H)$
          \State $wow\_pct\_abs \gets |\frac{f(t)-f(t-1 week)}{f(t-1 week)+1}|$
          \State $append(X, wow\_pct\_abs)$
          \State $H \gets H -1$
      \EndWhile
      \State $N \gets N - 1$
  \EndWhile
  \State $\tau = \overline{(X)} + 3\sigma(X)$ \Comment{Upper bound ratio}
  \State $N \gets 7$ \Comment{Get results only for the last week}
  \While{$N > 0$}
      \State $H \gets 24$ \Comment{Number of hours per day}
      \While {$H > 0$}
          \State $t \gets g(N, H)$ \Comment{an hour in the last week}
          \State $ub\_count = \lceil(\tau+1)(f(t)+1) - 1\rceil$
          \State $append(Y, (t, ub\_count))$
          \State $H \gets H -1$
      \EndWhile
      \State $N \gets N - 1$
  \EndWhile
  \State \textbf{Return:} Y
  \end{algorithmic}
  \end{algorithm}
  
We used Laplace Smoothing (Add-One Smoothing)~\cite{krishnan2020lipschitz} to avoid division by zero error.
We used the week-over-week change on the MAC-flap event count instead of the count itself because it helps to account for the natural variability in the data over time.
By calculating the week-over-week change, we are able to capture any significant increases or decreases in MAC-flap events over a short period of time.
This is important because the occurrence of MAC-flap events can vary significantly depending on the time-of-day, day-of-the-week, and other factors.
In addition, using the week-over-week change allows us to compare the current level of MAC-flap events to a more recent baseline, rather than comparing it to a long-term average.
This helps to capture any changes in MAC-flap events that may have occurred recently, which may be missed if we were only looking at long-term averages.
By setting the upper boundary for MAC-flap events based on three standard deviations from the mean week-over-week change, we are able to account for natural variability in the data while still capturing any significant changes in MAC-flap events that may indicate an anomaly.
This approach helps us to detect anomalies in MAC-flap events with a high degree of accuracy, while minimizing false positives.

Using the past $5$ weeks of data, we set the upper boundary for MAC-flap events based on three standard deviations ($3$-sigma) from the mean.
This upper boundary represents the expected range of MAC-flap event frequency for each network organization.
We then monitored the MAC-flap events for each organization and flagged any event that exceeded the upper boundary as an anomaly.

\begin{figure*}[!t]
  \centering
    \includegraphics[width=1.8\columnwidth]{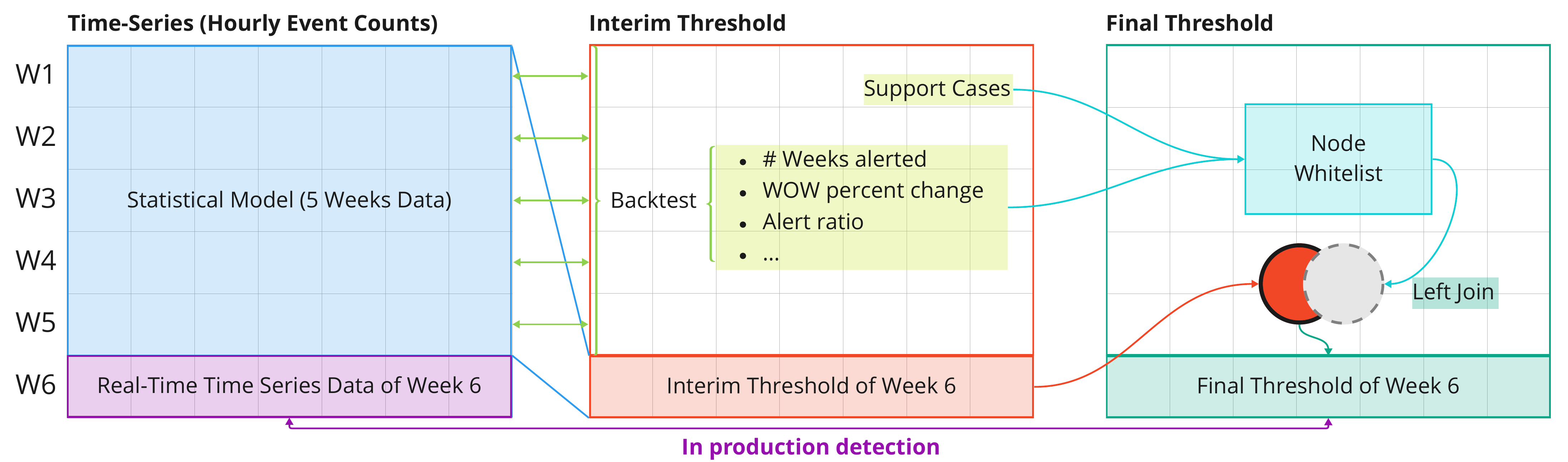}
  \vskip -.1in
  \caption{An example of how the threshold is generated on a weekly basis.}
  \label{fig:weekly_example}
  \vskip -.1in
\end{figure*}

\subsection{Whitelist Creation}
\label{sec:method_whitelist}

Enterprise-level products require not only accurate anomaly detection but also consideration of quality of service (QoS) and user experience.
It is important to minimize false positive alerts to prevent the alert system from generating unnecessary noise for customers.
In addition to the statistical model, a whitelist mechanism is included in our framework to filter out nodes that are persistently and stably alerted with a similar number of alerts week-over-week.
This additional mechanism provides greater control over the alert system and helps to reduce false positives.

In the specific use case of MAC-flap alerts, there are instances where MAC-flap events are expected and should not trigger an alert.
For example, when a wireless client is roaming from one access point (AP) to another (\textit{e.g.}, when someone is talking on the phone over Wi-Fi and walking between two APs in a cafeteria).
Therefore, it is essential to include direct control in the framework to filter out these expected events and prevent them from generating false alarms.

An example of the whole threshold generation workflow is depicted in Figure~\ref{fig:weekly_example}.
The statistical model described above in Section~\ref{sec:method_model} generates the interim thresholds for week $6$ (W6) based on previous $5$ weeks' time series data (W1 to W5).
At the same time, the previous $5$ weeks' time series data is back-tested with their corresponding interim thresholds to derive an alerting profile for each node, including \textit{i.e.} number of weeks alerted, week-over-week percent change of number of alerts or hourly event counts\footnote{More constraints and thresholds can be added when creating the whitelist.}.
These profiling results will be joined with support cases to determine a whitelist of nodes which will be exempted from being alerted -- if they persistently received alerts over $50\%$ of all the backtested weeks, yet no support case indicates that they are related to a connectivity issue.
Taking into account the interim threshold and the whitelist allows to effectively reduce false positive alerts.
Eventually, the final threshold of week $6$ will be deployed in production to stream alerts by comparing against real-time hourly event counts.

Overall, the combination of the statistical model and the whitelist mechanism provides a more robust and effective approach for anomaly detection in enterprise-level products. This approach not only ensures accurate detection but also takes into account the QoS and user experience, providing customers with a more reliable and satisfactory service.

\begin{figure}[!t]
  \centering
    \includegraphics[width=.7\columnwidth]{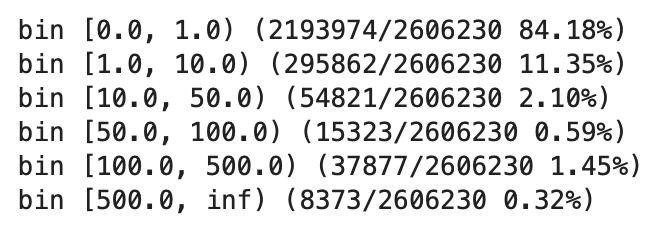}
  \vskip -.1in
  \caption{Histogram of MAC-flap events per hour.}
  \vskip -.1in
  \label{fig:eval_bin_distrib}
\end{figure}

\section{\uppercase{Evaluation}}

This section describes the model validation conducted on $20\%$ of switches (more than $117$k nodes) across $12$ weeks real-world data. 

\subsection{Model Performance}

In order to understand the distribution of MAC-flap events occurrence on network switches, we investigate the histogram of MAC-flap events per hour, as depicted in Figure~\ref{fig:eval_bin_distrib}.
$84.18\%$ nodes have no MAC-flap events and $11.35\%$ nodes have less than $10$ MAC-flap events per hour.
These two buckets account for more than $95\%$ of the whole population.

\begin{figure}[!t]
  \centering
    \includegraphics[width=.8\columnwidth]{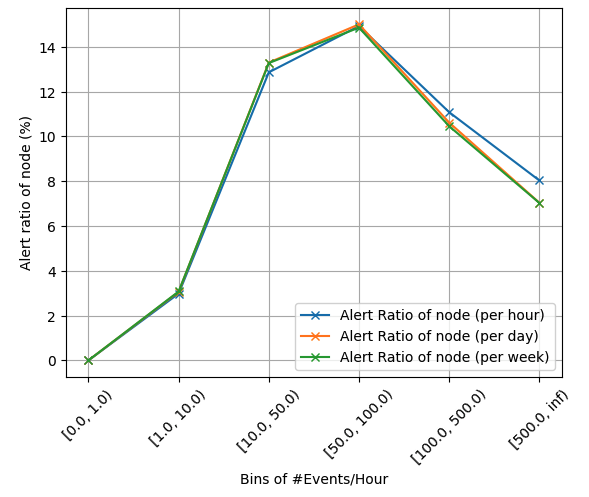}
    \vskip -.1in
    \caption{Alert ratio of nodes.}
  \label{fig:eval_alert_ratio_node}
  \vskip -.1in
\end{figure}

As depicted in Figure~\ref{fig:eval_alert_ratio_node}, the alert ratio of nodes based on the statistical model varies across the buckets and does not have linear dependency on the number of events. Network switches with $10$ to $100$ hourly MAC-flap events receive the highest ratio of alerts over time. However, nodes with more than $100$ hourly MAC-flap events are alerted less frequently. While intuitively networking devices with excessive amounts of MAC-flap events should be alerted, it is normal for some nodes to consistently observe flapping MAC addresses (e.g. roaming among APs as mentioned in Section~\ref{sec:method_whitelist}). Our model is able to differentiate nodes with persistently high amounts of hourly MAC-flap events.

\begin{figure}[!t]
  \centering
  \vskip -.1in
    \includegraphics[width=.8\columnwidth]{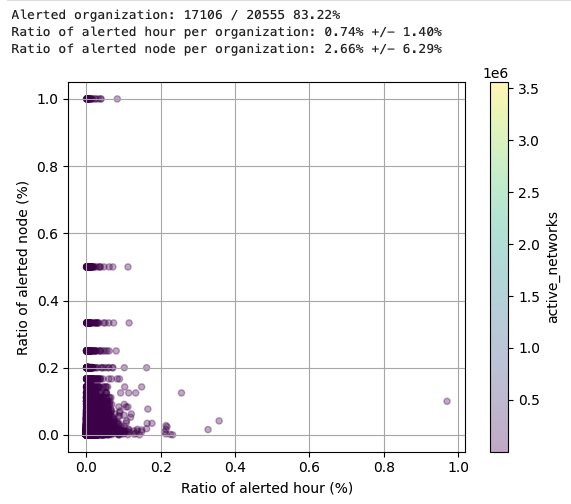}
  \vskip -.1in
  \caption{Alert ratio of organizations.}
  \label{fig:eval_alert_ratio_org}
  \vskip -.1in
\end{figure}

Figure~\ref{fig:eval_alert_ratio_org} segments the generated alerts by organizations (each organization may have thousands of nodes) and depicts on $2$ axes -- \textit{i.e.} the ratio of alerted hours and nodes. While ratio of alerted organizations (at least alerted once) is higher than $80\%$, the alerted ratios of hours and nodes per organization are low. Most organizations locate at the bottom left, with low alert ratio on both dimensions. In the figure, the top-left region represents the organizations that have a group of nodes alerted during a brief period of time, while the bottom-right region represents the organizations that have a small subset of nodes persistently alerted over time.

\begin{figure}[!t]
  \centering
    \includegraphics[width=.8\columnwidth]{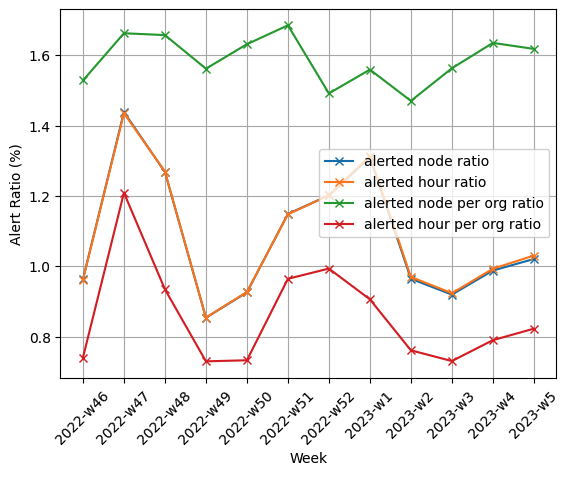}
  \vskip -.1in
  \caption{Alert ratio across $12$ weeks.}
  \label{fig:eval_wow_alert}
  \vskip -.1in
\end{figure}

As the thresholds are generated on a weekly basis, we compared the alert ratios across the $12$ weeks. Figure~\ref{fig:eval_wow_alert} demonstrates that the ratios of alerts generated by the statistical model are stable over weeks.

\subsection{Model Validation w/ Support Cases}
\label{sec:eval_case}

The alerts are cross-validated with customer support cases related to MAC-flap events signaling connectivity issues they encountered. MAC-flap events happened before and after $30$ days to when the customer support cases are opened are considered as ground truth labeled data\footnote{The delay is derived from an estimation of time for the customer to discover networking issues and to be ensured that the issues are entirely resolved.}.

\begin{figure}[!t]
  \centering
    \includegraphics[width=.8\columnwidth]{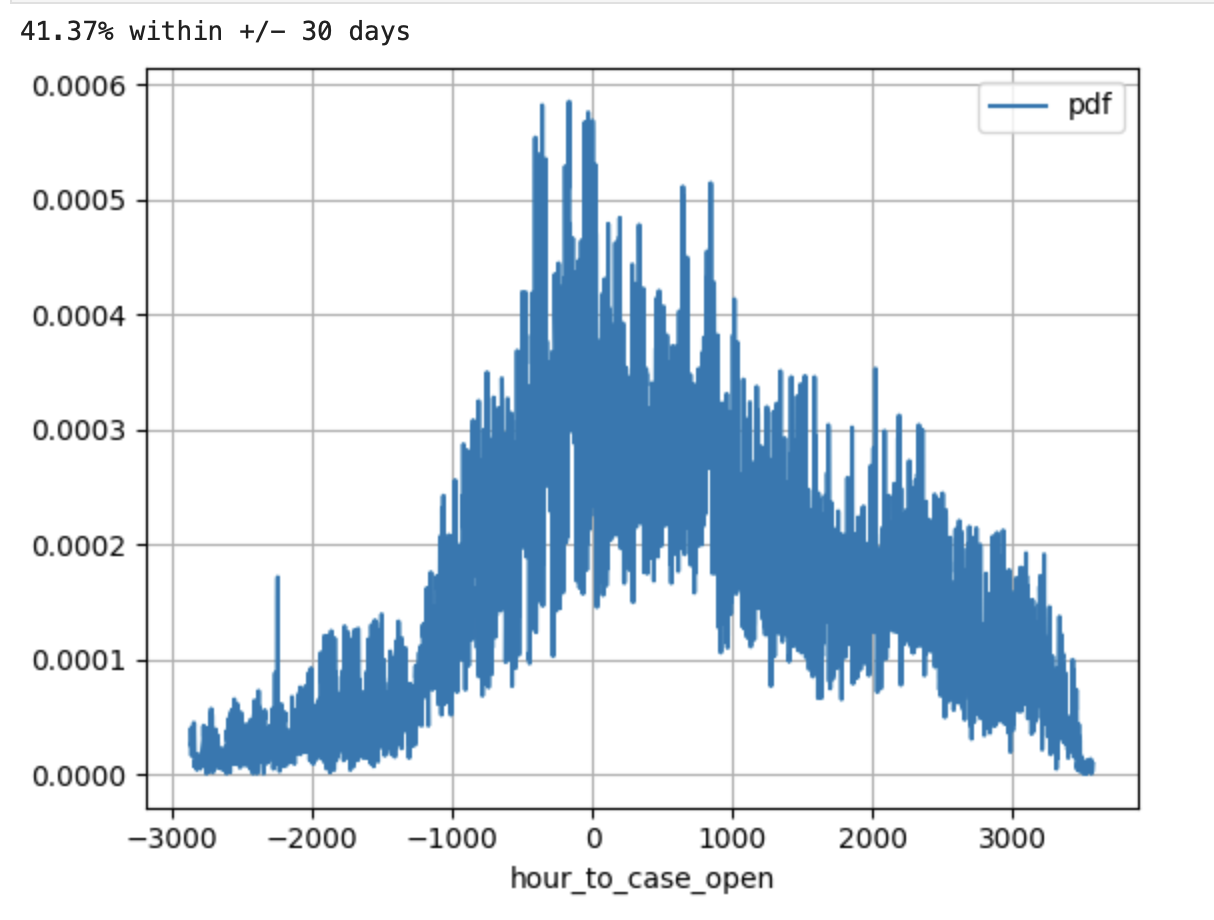}
  \vskip -.1in
  \caption{MAC-flap events cross-validated with support cases.}
  \label{fig:eval_case_valid}
  \vskip -.1in
\end{figure}

As depicted in Figure~\ref{fig:eval_case_valid}, $41.37\%$ of hourly MAC-flap event counts fall within the $30$-day ($720$-hour) time range before and after the case opening. The distribution of the temporal distance between MAC-flap events and support cases indicates that these events are potentially subject to connectivity issues.

\begin{figure}[!t]
  \centering
    \includegraphics[width=.8\columnwidth]{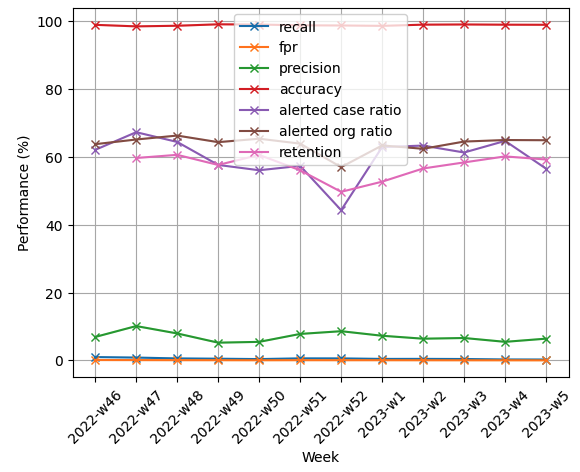}
  \vskip -.1in
  \caption{Model performance across $12$ weeks.}
  \label{fig:eval_wow_metric}
  \vskip -.1in
\end{figure}

Across $12$ weeks, the statistical model performance is depicted in Figure~\ref{fig:eval_wow_metric}, which also shows stability over weeks. High accuracy is achieved by predicting most true negatives (TN) ($99.91\%$ TN rate).
The output of the statistical model successfully alerts more than $60\%$ of support cases, yet at the cost of alerting more than half of organizations and persistently alerting a subset of nodes with no relation to the support cases (the line of \textit{retention} in the figure), which leads to low precision and recall.
These findings justified the requirement of including a whitelist to reduce false positives (FP), and decrease the ratio of alerted organizations, therefore make the alert system less noisy.

\begin{figure}[!t]
  \centering
    \includegraphics[width=.8\columnwidth]{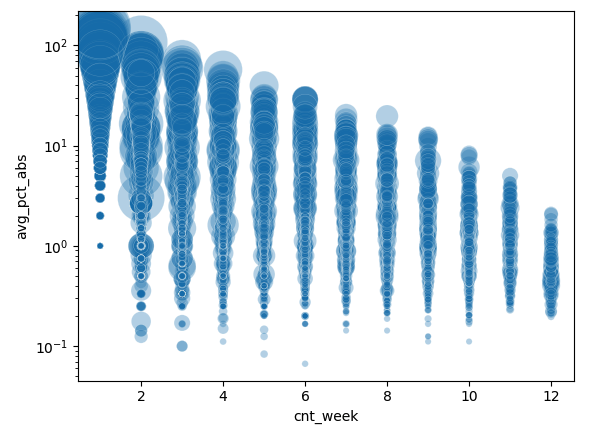}
  \vskip -.1in
  \caption{Whitelist creation based on $2$ thresholds (x- and y-axes). The bubble size represents the number of alerts per week a node receives.}
  \label{fig:eval_whitelist}
  \vskip -.1in
\end{figure}

Based on the previous observations, a whitelist can be created based on the back-testing procedure described in Section~\ref{sec:method_whitelist}. It aims for excluding network switches that are persistently alerted over weeks with stable amount of alerts each week, which are the bubbles located at the bottom-right region in Figure~\ref{fig:eval_whitelist}. For instance, it helps to reduce $32.3\%$ of nodes from being alerted by configuring the thresholds to exclude nodes that are alerted more than $6$ weeks across $12$ weeks, and have the average of absolute week-over-week percent change of number of alerts lower than $10$.

\begin{figure}[!t]
  \centering
    \includegraphics[width=\columnwidth]{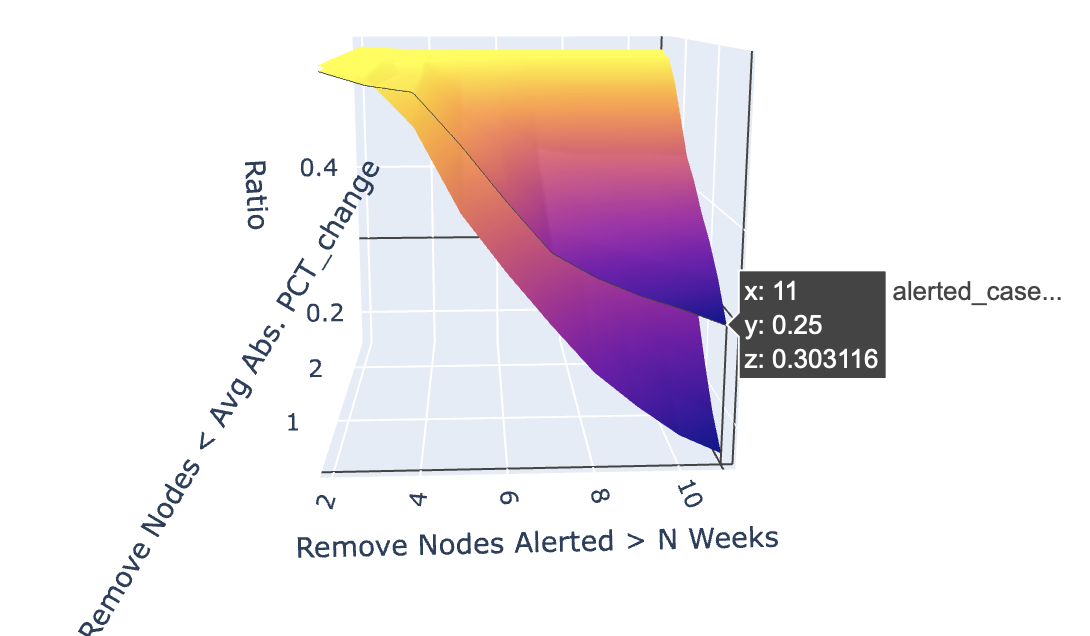}
  \caption{Tuning alert ratio by adjusting the $2$ thresholds to achieve a balanced trade-off between alerted organizations and alerted case-related data points.}
  \label{fig:eval_3d}
\end{figure}

As depicted in Figure~\ref{fig:eval_3d}, conducting grid search over the $2$ thresholds allows to largely decrease the ratio of alerted organizations from more than $60\%$ to less than $20\%$ while alerting more than $30\%$ of MAC-flap-related support cases.

\subsection{Discussion}

This study presents an initial attempt towards developing a hybrid model for a more effective alert system by integrating statistical models, backtesting, and qualitative data validation. The hybrid model aims to enhance the performance of the existing alert system by reducing false positive alerts through a two-step approach, i.e., employing statistical models and utilizing backtesting over multiple weeks to filter out persistently alerted nodes. Furthermore, the model's effectiveness is validated through the analysis of qualitative data gathered from customer support cases.

Subsequent studies can focus on improving the proposed model's performance by expanding the input data set and conducting further feature engineering. The study suggests exploring correlations between events such as MAC-flap and layer-2 loop detection to identify underlying patterns that can aid in the development of more accurate alert systems.
Additionally, future research may consider including more human feedback than the support cases as a part of the model development process to improve the user experience. For instance, the study proposes A/B testing to gather customer feedback and ensure the usefulness of the new alert system in accurately predicting network connectivity issues.

\section{\uppercase{Conclusions}}
\label{sec:conclusion}

In this paper, we proposed a hybrid model for an alert system that combines statistical models with a whitelist mechanism to reduce false positive alerts in network management. Our approach leverages the large database to detect anomalies and incorporates a whitelist mechanism to filter out persistently alerted nodes, resulting in a more accurate and efficient alert system. We have validated our model using qualitative data from customer support cases and identified opportunities for future work, including more feature engineering and input data, as well as incorporating more human feedback in the model development process. Our proposed approach offers a promising solution to the challenge of reducing false positive alerts in network management and has the potential to improve the user experience in enterprise-level products.

\section*{\uppercase{Acknowledgement}}

We would like to express our sincere gratitude to our data engineers -- Rajat Mehrotra, Sinduja Seethapathy, Vinay Kumar Abburi, and Nate Fung, for obtaining the data for model validation, maintaining the data infrastructure, and checking the data quality.
We greatly appreciate their invaluable support and dedication to this project.

\bibliographystyle{IEEEtran}
\bibliography{reference}

\end{document}